\documentclass[reprint,aps,prl,superscriptaddress,amsmath,amssymb,floatfix]{revtex4-2}

\usepackage{graphicx,color}
\usepackage{dcolumn}
\usepackage{bm}
\usepackage{comment}
\usepackage{float}

\usepackage{textcase, url}
\usepackage[normalem]{ulem}
\usepackage{siunitx}

\usepackage{mathcomp} 

\definecolor{darkblue}{rgb}{0,0,0.5}
\definecolor{darkgreen}{rgb}{0,0.5,0}
\definecolor{darkred}{rgb}{.7,0,0}
\definecolor{purple}{rgb}{0.5,0,0.6}
\definecolor{orange}{rgb}{1,0.5,0}
\definecolor{grey}{rgb}{.6,.6,.6}
\definecolor{lightpink}{rgb}{1,0.7,0.75}
\definecolor{pink}{rgb}{1,0.4,0.58}
\definecolor{deeppink}{rgb}{1,0.08,0.58}
\definecolor{brown}{rgb}{0.59, 0.29, 0.0}
\definecolor{blue-green}{rgb}{0.0, 0.87, 0.87}

\begin{document}


\title{Observation of Shapiro Steps in the Charge Density Wave State \\ 
Induced by Strain on a Piezoelectric Substrate}

\author{Koji~Fujiwara}
\affiliation{
 Department of Physics, Graduate School of Science, Osaka University, Osaka 560-0043, Japan
 }
\author{Takuya~Kawada}%
\affiliation{
 Department of Physics, Graduate School of Science, Osaka University, Osaka 560-0043, Japan
 }
 \affiliation{
 Department of Basic Science, The University of Tokyo, Tokyo 153-8902, Japan
 }
\author{Natsumi~Nikaido}
\affiliation{
 Department of Physics, Graduate School of Science, Osaka University, Osaka 560-0043, Japan
 }
\author{Jihoon~Park}
\affiliation{
 Department of Physics, Graduate School of Science, Osaka University, Osaka 560-0043, Japan
 }
\author{Nan~Jiang}
\affiliation{
 Department of Physics, Graduate School of Science, Osaka University, Osaka 560-0043, Japan
 }
\affiliation{
 Center for Spintronics Research Network, Osaka University, Osaka 560-8531, Japan
 }
\affiliation{
 Institute for Open and Transdisciplinary Research Initiatives, Osaka University, Osaka 565-0871, Japan
 }
\author{Shintaro~Takada}
\affiliation{
 Department of Physics, Graduate School of Science, Osaka University, Osaka 560-0043, Japan
 }
\affiliation{
 Center for Spintronics Research Network, Osaka University, Osaka 560-8531, Japan
 }
\affiliation{
 Institute for Open and Transdisciplinary Research Initiatives, Osaka University, Osaka 565-0871, Japan
 }
\author{Yasuhiro~Niimi}
\email{niimi@phys.sci.osaka-u.ac.jp}
\affiliation{
 Department of Physics, Graduate School of Science, Osaka University, Osaka 560-0043, Japan
 }
\affiliation{
 Center for Spintronics Research Network, Osaka University, Osaka 560-8531, Japan
 }
\affiliation{
 Institute for Open and Transdisciplinary Research Initiatives, Osaka University, Osaka 565-0871, Japan
 }

\date{\today}
\begin{abstract}
Recent development in nanotechnology has enabled us to 
investigate the dynamic properties of van der Waals materials 
on a piezoelectric substrate.
Here we report on the dynamics of charge density wave (CDW) 
in NbSe$_{3}$ nanowires induced by surface acoustic waves (SAWs). 
Clear peaks in the differential resistance were observed 
at the resonant frequency of the SAW device. 
These peaks known as Shapiro steps are typically observed 
by applying an rf current to NbSe$_{3}$ nanowires. 
We found that the Shapiro steps induced by SAWs show several distinct features 
from the ones induced by an rf current. Our detailed study revealed 
that a strain induced by SAWs plays a significant role in the Shapiro steps. 
The result clearly demonstrates the importance of the strain in CDW materials 
and paves the way for strain-induced device applications.
\end{abstract}

\maketitle

Strain and distortion are fundamental concepts in physics. 
In geology, for example, strain refers to the deformation or change in shape 
and size of a solid material due to applied stress. 
The occurrence of earthquakes can be attributed to the release of accumulated strain~\cite{lawson_1908_california}. 
In general relativity, gravity arises as a result of the distortion of spacetime~\cite{Einstein_1916}. 
In condensed matter physics, on the other hand, strain can be introduced artificially 
by applying pressure to a material or bending a material. 
In addition, it can be controlled electrically by applying an electric field to a 
piezoelectric material. 
This principle underlies surface acoustic waves (SAWs)
-- sound waves that propagate along the surface of a solid.
SAWs can be excited by applying an ac voltage
to an interdigital transducer (IDT)
fabricated on a piezoelectric substrate~\cite{White_1965, Delsing_2019}.

Recently, SAWs have been utilized to modulate the electronic properties of 
two-dimensional (2D) electronic systems: not only  
in a 2D electron gas embedded in piezoelectric GaAs/AlGaAs heterojunctions~\cite{Hermelin_nat_2011, McNeil_nat_2011, Bertrand_natnano_2016, Takada_natcom_2019, Edlbauer_apl_2021, Wang_prx_2022}
but also a variety of 2D materials such as graphene and transition metal 
dichalcogenides placed on a piezoelectric LiNbO$_3$ 
substrate~\cite{Miseikis_2012, Bandhu_2014, Preciado_2015, Nie_2023, Zhao_2024, Fandan_2020, Zhao_2022, Fang_2023, Lyons_2023, yokoi_sci_adv_2020}. 
In particular, in the latter case, the electronic properties of 2D materials 
on the substrate have been modulated by SAWs 
through electron-phonon couplings and magnetoelastic couplings, 
owing to the large piezoelectric constant of LiNbO$_3$. 


Here we focus on charge density waves (CDWs)~\cite{Peierls_1955}.
In low-dimensional conductors with strong electron-lattice interaction, 
a lattice distortion occurs below a certain transition temperature,
leading to a modulation of the electron density with the same wavenumber
as the lattice distortion. 
In the CDW state, an energy gap opens partially at the Fermi level,
resulting in an increase of resistance.
When an electric field applied to the CDW exceeds a threshold value,
the CDW begins to slide, which causes a decrease in resistance~\cite{gruner_rmp_1988}.
From the theoretical viewpoint, the motion of the CDW
can be modeled as a point mass in a washboard potential, where
the CDW is periodically accelerated and decelerated.
This motion is analogous to the behavior of a Josephson junction~\cite{Shapiro_1963},
and Shapiro steps are also observed in CDW materials.
So far, such Shapiro steps in the CDW state have been observed 
by directly applying an rf 
current~\cite{Monceau_1980, Zettl_1983, Zettl_1984, Thorne_1988}
or by mechanically driving the CDW material at radio frequencies~\cite{Nikitin_APL_2021}. 
Although a strain in a CDW material should give a strong impact on the CDW dynamics, 
the relation between the strain and the Shapiro steps have not been 
elucidated yet. 
The aforementioned thin-film device on a piezoelectric substrate 
offers a promising approach to address the issue.

In this Letter, we have investigated 
the CDW dynamics of mechanically exfoliated NbSe$_3$ nanowires 
on a piezoelectric LiNbO$_{3}$ substrate on which SAWs with 
a frequency of sub-Giga Hertz are generated. 
Several peaks corresponding to the Shapiro steps appear 
in the differential resistance only at the resonant frequency of the SAW. 
The widths of the Shapiro steps oscillate as a function of 
the SAW power, as in the case of the rf current. 
However, the oscillations observed with the SAW devices 
decay much faster than those with the conventional setup. 
Combined with numerical simulations, we found that the Shapiro steps 
on the piezoelectric substrate originate not from the electric field 
induced by the substrate but from the strain of the substrate.

\begin{figure}
\centering
\includegraphics[width=85mm]{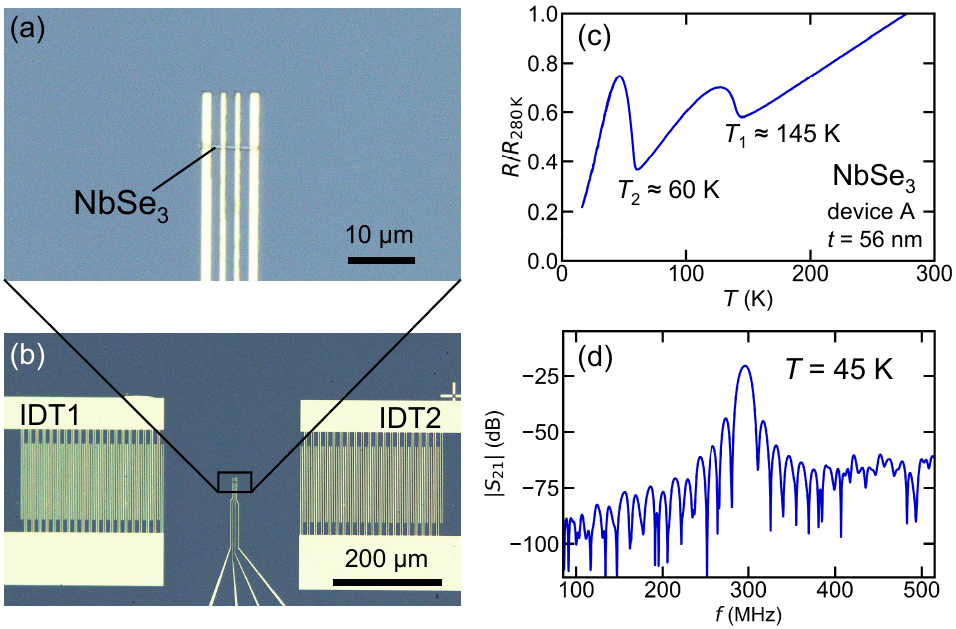}
\caption{\label{figure1} (a), (b) Optical microscope images of (a) an exfoliated NbSe$_{3}$ nanowire and (b) the whole SAW device. The NbSe$_{3}$ nanowire is allocated in between two interdigital transducers (IDTs), i.e., IDT1 and IDT2. (c) Temperature dependence of resistance $R$ of the NbSe$_{3}$ nanowire with a thickness $t = \SI{56}{nm}$ (device A), normalized by $R$ at $T = \SI{280}{K}$. Two bumps due to the charge density wave (CDW) transitions are clearly observed. (d) Scattering parameter from IDT1 to IDT2 ($|S_{21}|$) on a logarithmic scale as a function of the frequency $f$ measured at $T = \SI{45}{K}$. This is the result after the time domain gating process is applied to remove electromagnetic crosstalk signals. The transmitted signal intensity has a peak at $f_{0} = \SI{296}{MHz}$.}
\end{figure}

NbSe$_{3}$ is one of the typical quasi-one-dimensional crystals and 
known to show two incommensurate CDW transitions~\cite{tsutsumi_prl_1977,fleming_prb_1978,ong_prb_1977,thorne_prl_2001}.
One is the linear nesting along the $b$-axis below $T_{1} \approx \SI{145}{K}$ 
which we call CDW1. 
The other is the diagonal nesting (along the $a$-$c$ plane as well as the $b$-axis) 
below $T_{2} \approx \SI{60}{K}$ which we call CDW2.
Below $T_{2}$, both CDW1 and CDW2 are realized at the same time.
NbSe$_{3}$ nanowires on a piezoelectric 128$^\circ$Y-cut black LiNbO$_{3}$ substrate 
were obtained through the mechanical exfoliation technique using scotch tapes 
inside a glovebox filled with Ar gas. 
As shown in Figs.~\ref{figure1}(a) and \ref{figure1}(b), 
electrodes for the NbSe$_{3}$ nanowire and a pair of IDTs for the excitation of SAW 
on the substrate were patterned by means of electron beam lithography and 
obtained by depositing Ti(\SI{30}{nm})/Au(\SI{40}{nm}) and 
performing the subsequent liftoff process. 
In this work, we prepared two types of SAW devices: one has 
a resonant frequency $f_{0} \approx \SI{300}{MHz}$ 
and the other does $f_{0} \approx \SI{2}{GHz}$. 
We mainly show the results with the former type of device because 
the electromagnetic waves from the IDTs can significantly affect the results 
as shown in Fig.~S1 in the Supplemental Material~\cite{supplement}.

In Fig.~\ref{figure1}(c), we show a resistance $R$ versus temperature $T$ 
curve for a typical NbSe$_{3}$ nanowire.  
Two bumps due to the CDW transitions 
are clearly seen at $T_{1} \approx \SI{145}{K}$ and $T_{2} \approx \SI{60}{K}$, 
as in the case of bulk NbSe$_{3}$~\cite{ong_prb_1977} and NbSe$_3$ 
nanowire~\cite{onishi_njp_2017,stabile_nt_2011,hor_nanolett_2005,Yang_2019,Fujiwara_jjap_2021}. 
To check basic properties of IDTs,
we measured an $S$-parameter from one of the IDTs (IDT1) 
to the other (IDT2), 
that is $|S_{21}|$, at $T = \SI{45}{K}$ in Fig.~\ref{figure1}(d).
The intensity of the transmitted signal takes a maximum
at the resonant frequency of $f_{0} = \SI{296}{MHz}$, which 
is consistent with the expected value $f_{0} = v_{0}/\lambda$ 
where $v_{0} \approx \SI{3900}{m/s}$ is 
the sound velocity along the crystalline X-axis on the 128$^\circ$Y-cut LiNbO$_{3}$ substrate 
and $\lambda = \SI{13.2}{\um}$ is
the period of IDT fingers.

\begin{figure}
\centering
\includegraphics[width=65mm]{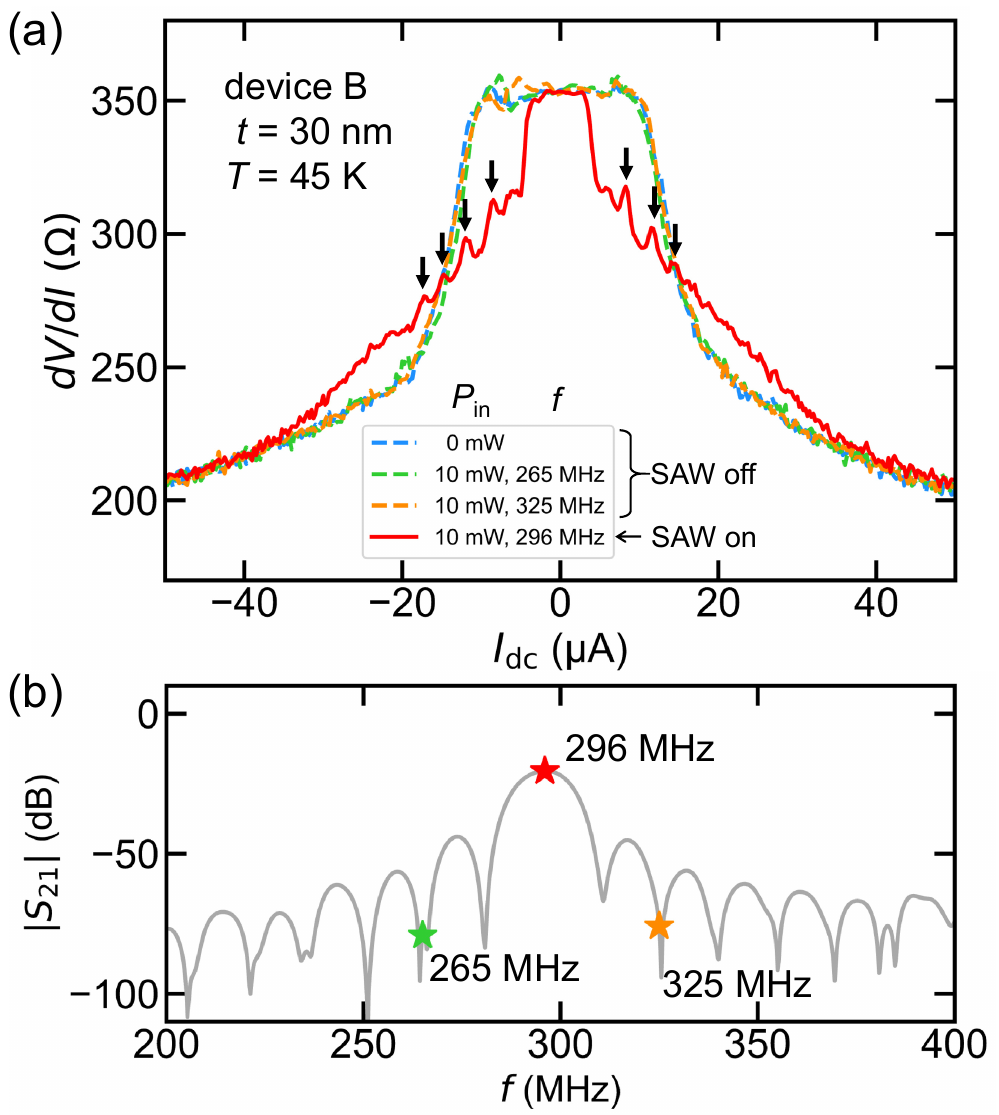}
\caption{\label{figure2} Differential resistance $dV/dI$ as a function of direct current $I_{\mathrm{dc}}$ measured at $T = \SI{45}{K}$ with different frequencies. $P_{\mathrm{in}}$ is the power applied to IDT1 and fixed at \SI{10}{mW} in this measurement.
(b) Scattering parameter $|S_{21}|$ as a function of the frequency $f$ measured at $T=\SI{45}{K}$. The stars in the figure indicate the measured frequencies in (a).}
\end{figure}

In order to focus on the CDW dynamics, 
we plot a differential resistance $dV/dI$ at $T = \SI{45}{K}$ (below $T_{2}$)
as a function of direct current $I_{\mathrm{dc}}$ 
applied to the nanowire in Fig.~\ref{figure2}(a). 
When there is no SAW excitation 
(i.e., the power applied to the IDTs $P_{\mathrm{in}}$ is zero), 
$dV/dI$ is constant up to a threshold value and 
starts to decrease at $I_{\mathrm{dc}} \approx \SI{10}{\uA}$ 
[see the light blue broken line in Fig.~\ref{figure2}(a)]
because of the CDW sliding as mentioned in the introduction. 
On the other hand, when the SAW is irradiated to the nanowire 
through the IDTs ($P_{\mathrm{in}} = \SI{10}{mW}$ at $f = f_{0}= \SI{296}{MHz}$), 
several peaks indicated by the arrows in Fig.~\ref{figure2}(a) 
appear in the $dV/dI$ vs $I_{\mathrm{dc}}$ curve. 
These peaks correspond to the Shapiro steps, 
which can be seen with the application of an rf current to 
CDW compounds~\cite{Monceau_1980, Zettl_1983, Zettl_1984, Thorne_1988}. 
The difference between the present SAW device and the conventional rf setup 
is that the Shapiro steps appear only at the resonant frequency of the IDTs. 
As shown in Fig.~\ref{figure2}(b), when some off-resonant frequencies 
are selected, no peaks are observed in the $dV/dI$ vs $I_{\mathrm{dc}}$ 
curves [see the light green and orange broken lines and symbols in Fig.~\ref{figure2}].
This indicates that the Shapiro steps are induced by the SAWs
rather than by electromagnetic waves emitted from the IDTs.

We note that $P_{\mathrm{in}}$ is not the SAW power ($P_\mathrm{SAW}$) 
that is applied to the NbSe$_{3}$ nanowire. 
$P_\mathrm{SAW}$ can be converted from 
$P_{\mathrm{in}}$ using a measured $|S_{21}|$ value~\cite{Fandan_2020} as shown below:
\begin{equation*}
     P_{\mathrm{SAW}}\:[\mathrm{W}] = 10^\frac{|S_{21}| / 2 \: [\mathrm{dB}]}{10} P_{\mathrm{in}} \:[\mathrm{W}].   
\end{equation*}
For example, $P_{\mathrm{in}} = \SI{10}{mW}$ at $f = f_{0} = \SI{296}{MHz}$ 
corresponds to $P_{\mathrm{SAW}} \approx \SI{0.95}{mW}$ [see Fig.~\ref{figure2}]. 
Hereafter, we use $P_{\mathrm{SAW}}$ instead of $P_{\mathrm{in}}$.

\begin{figure}
\centering
\includegraphics[width=80mm]{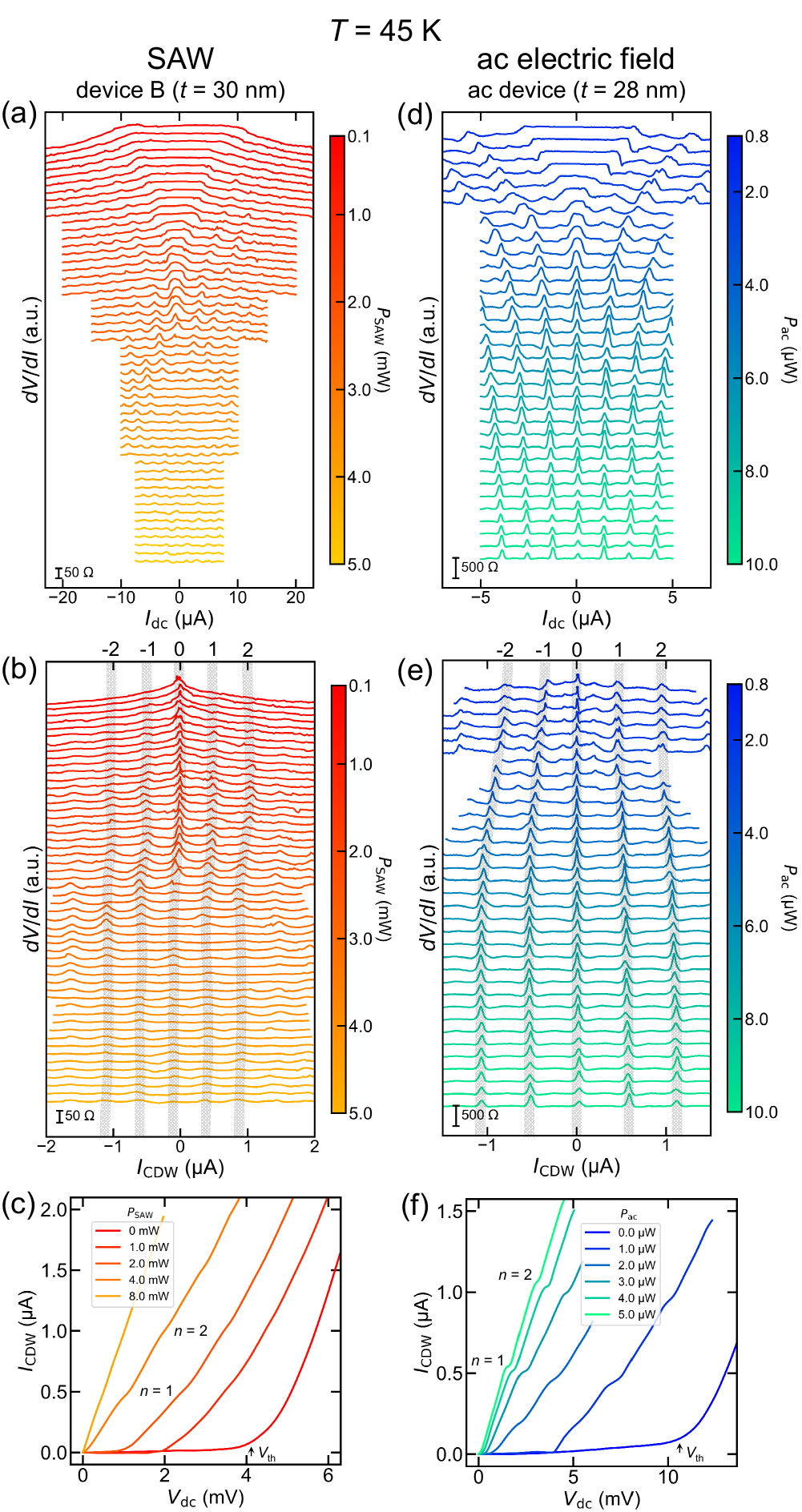}
\caption{\label{figure3_45K} (a) $dV/dI$ as a function of $I_{\mathrm{dc}}$ measured at $T=\SI{45}{K}$ with different $P_{\mathrm{SAW}}$. (b) $dV/dI$ as a function of $I_{\mathrm{CDW}} \equiv I_{\mathrm{dc}} - I_{\mathrm{ohmic}}$ at $T = \SI{45}{K}$ with different $P_{\mathrm{SAW}}$. (c) $I_{\mathrm{CDW}}$ as a function of dc voltage $V_{\mathrm{dc}}$, obtained by integrating the $dV/dI$ curve, at typical $P_{\mathrm{SAW}}$ values. (d) $dV/dI$ as a function of $I_{\mathrm{dc}}$ measured at $T = \SI{45}{K}$ with different ac powers $P_{\mathrm{ac}}$. (e) $dV/dI$ as a function of $I_{\mathrm{CDW}}$ at $T = \SI{45}{K}$ with different $P_{\mathrm{ac}}$. (f) $I_{\mathrm{CDW}}$ as a function of dc voltage $V_{\mathrm{dc}}$ at typical $P_{\mathrm{ac}}$ values. The steps for $n=1$ and $n=2$ are indicated in (c) and (f). The threshold voltage $V_{\mathrm{th}}$ to drive CDW is also defined in (c) and (f).}
\end{figure}

We then measured the $dV/dI$ vs $I_{\mathrm{dc}}$ curve at $f_{0}$ with 
different SAW powers at \SI{45}{K} as shown in Fig.~\ref{figure3_45K}(a). 
With increasing $P_{\mathrm{SAW}}$, we observed several peaks 
originating from the Shapiro steps. 
To characterize the Shapiro steps, 
we first subtracted the normal electrical conduction $I_{\mathrm{ohmic}}$
(i.e., the plateau in the $dV/dI$ vs $I_{\mathrm{dc}}$ curve) and 
plotted $dV/dI$ as a function of 
$I_{\mathrm{CDW}} \equiv I_{\mathrm{dc}} - I_{\mathrm{ohmic}}$, 
as depicted in Fig.~\ref{figure3_45K}(b). 
From this graph, we can identify the index number $n$ of the Shapiro steps. 
While the peak at $I_{\mathrm{CDW}} = 0$ originates from the CDW sliding (i.e., $n=0$), 
a couple of peaks appear at almost the same $I_{\mathrm{CDW}}$ values 
with increasing $P_{\mathrm{SAW}}$, as indicated gray lines. 
These correspond to the Shapiro steps. 
In order to see the steps more clearly~\cite{supplement}, 
we integrated the $dV/dI$ vs $I_{\mathrm{dc}}$ curve 
(resulting in $V_{\mathrm{dc}}$ vs $I_{\mathrm{dc}}$ curve)  
and plotted $I_{\mathrm{CDW}}$ as a function of $V_{\mathrm{dc}}$ 
at different $P_{\mathrm{SAW}}$ values in Fig.~\ref{figure3_45K}(c). 
$I_{\mathrm{CDW}}$ is zero up to the threshold voltage $V_{\mathrm{th}}$, 
and starts to increase above $V_{\mathrm{th}}$. 
Plateau-like structures can be seen at $I_{\mathrm{CDW}} \approx \SI{0.5}{\uA}$ and 
$\approx \SI{1}{\uA}$, which correspond to 
the Shapiro steps $n=1$ and $n=2$, respectively.

It is well-established that the Shapiro steps in CDW materials can be 
observed by directly applying an rf current to 
the samples~\cite{Monceau_1980, Zettl_1983, Zettl_1984, Thorne_1988}. 
To compare the Shapiro steps induced by SAW with those obtained with the conventional setup, 
we prepared another NbSe$_{3}$ nanowire with almost the same thickness ($t = \SI{28}{nm}$)
on a Si substrate, and measured Shapiro steps 
by directly flowing an alternating current with $f = \SI{300}{MHz}$ to the nanowire. 
In Fig.~\ref{figure3_45K}(d), 
we show $dV/dI$ as a function of $I_\mathrm{dc}$
at different $P_{\mathrm{ac}}$ values.
By performing the same analysis as for Figs.~\ref{figure3_45K}(b) and \ref{figure3_45K}(c), 
the relations between $dV/dI$ and $I_{\mathrm{CDW}}$ and between 
$I_{\mathrm{CDW}}$ and $V_{\mathrm{dc}}$ are obtained as shown in 
Figs.~\ref{figure3_45K}(e) and \ref{figure3_45K}(f), respectively.

\begin{figure*}
    \centering
    \includegraphics[width=165mm]{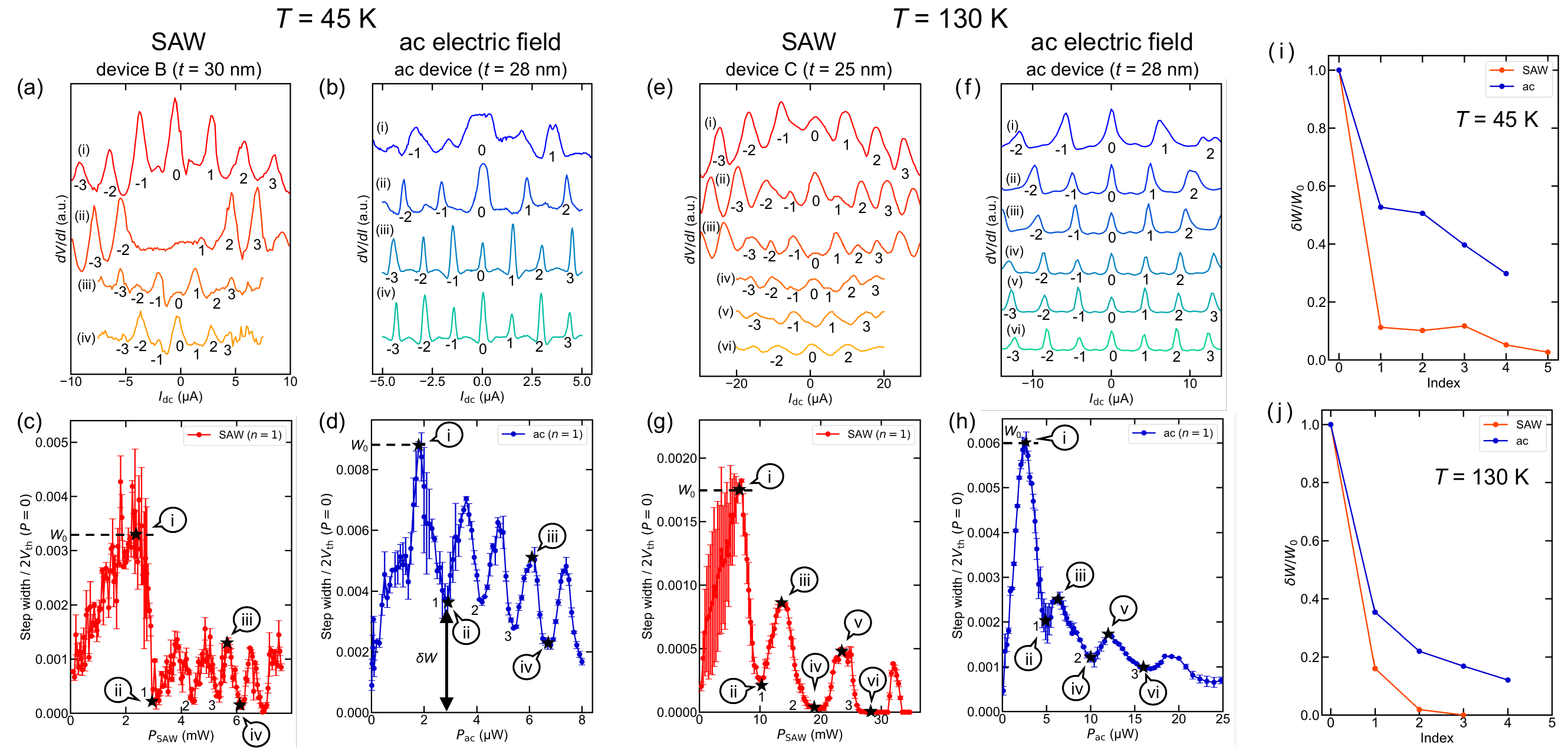}
    \caption{(a), (b) $dV/dI$ as a function of $I_{\mathrm{dc}}$ measured with (a) the SAW device (device B) and (b) the ac device at $T = \SI{45}{K}$. We present some representative power values where the step width of the $n = 1$ step takes a maximum and minimum. (c), (d) The step widths divided by $2V_{\mathrm{th}}(P=0)$ as a function of (c) $P_{\mathrm{SAW}}$ and (d) $P_{\mathrm{ac}}$ for the $n=1$ step. For both devices, the thicknesses $t$ of NbSe$_3$ thin films are about \SI{30}{nm}. $W_{0}$ is the step width maximum and $\delta W$ is the step width at each minimum point. The indices of the minimum points are also defined in the figures. (e)--(h) The same datasets obtained at $T = \SI{130}{K}$. (i), (j) $\delta W/W_{0}$ as a function of the index number obtained at (i) $T = \SI{45}{K}$ and (j) $T = \SI{130}{K}$.}
    \label{fig:dVdI_pickup}
\end{figure*}

Now, we compare the two cases. 
Figures~\ref{fig:dVdI_pickup}(a) and \ref{fig:dVdI_pickup}(b) show 
the $dV/dI$ vs $I_{\mathrm{dc}}$ curves obtained with the SAW device 
and the conventional ac device, respectively, at typical $P_{\mathrm{SAW}}$ 
and $P_{\mathrm{ac}}$ values at $T=\SI{45}{K}$.
For both cases, the amplitude of the $dV/dI$ peak changes with the power. 
It should be noted that some of the peaks are almost invisible for the SAW device 
[see, for example, $n=1$ at (ii) and (iv), $n=2$ at (iii) in Fig.~\ref{fig:dVdI_pickup}(a)], 
while all the peaks still remain distinct for the ac device. 
This situation can be clearly seen in Figs.~\ref{fig:dVdI_pickup}(c) 
and \ref{fig:dVdI_pickup}(d) where the width of the 
Shapiro step divided by $2V_{\mathrm{th}}~(P = 0)$ 
for $n=1$ is plotted as a function of 
$P_{\mathrm{SAW}}$ and $P_{\mathrm{ac}}$, respectively.
The step width is defined as the integrated area of the peak
in $dV/dI$ vs $I_\mathrm{dc}$ plot~\cite{BROWN_1989_p223}.
In the conventional ac device, the step width oscillates 
as a function of $P_{\mathrm{ac}}$ and can be described by Bessel function 
[see Fig.~\ref{fig:dVdI_pickup}(d)]. 
This is the same tendency as the previous 
studies~\cite{Zettl_1983, Zettl_1984, Thorne_1988, Zybtsev_2017, Zybtsev_2020}.
For the case of the SAW device, the step width also oscillates similar
to the conventional steps [see Fig.~\ref{fig:dVdI_pickup}(c)]. 
This behavior is different from the theoretical calculations~\cite{Funami_PRB_2023}
where the step width monotonically increases with increasing a pinning-strength 
modulation in the model. 
On the other hand, the SAW power dependence of the step width is 
apparently different from the conventional one: 
the step width oscillates but with a much faster reduction than the case 
of the conventional ac device and almost vanishes at some $P_{\mathrm{SAW}}$ values.
The similar tendency has been confirmed 
at different peaks $n=\pm 2$~\cite{supplement}, 
and for the other CDW (CDW1) at $T=\SI{130}{K}$ 
[see Figs.~\ref{fig:dVdI_pickup}(e)--\ref{fig:dVdI_pickup}(h)], 
and also using four other NbSe$_{3}$ devices (see Fig.~S4 in ~\cite{supplement}).

In order to see the faster damping of the step width for the SAW device more clearly, 
we plot the step width at each minimum point $\delta W$ 
normalized by the maximum value $W_0$ 
as a function of the minimum index number as shown 
in Figs.~\ref{fig:dVdI_pickup}(i) and \ref{fig:dVdI_pickup}(j). 
Compared to the case of the ac device, 
$\delta W/W_{0}$ of the SAW device decreases more rapidly 
with increasing the index for both the temperatures \SI{45}{K} and \SI{130}{K}.

Based on the washboard model, the step width drops completely to zero. 
This is because the washboard model assumes that there is no spatial variation 
in the phase of the CDW. 
However, this assumption is not realistic because 
there are many impurities, defects, grain boundaries in the CDW material.
In such a case, the step width generally does not drop to zero,
as in the case of the ac device, but takes a finite value 
even at a minimum point~\cite{Funami_PRB_2023}.
On the other hand, in the case of the SAW device, 
the step width not only drops to almost zero
but also is significantly reduced with increasing the SAW power,
compared to the case of the ac device.

While the two situations are qualitatively different, 
there are a few things to be addressed. 
$P_{\mathrm{ac}}$ is directly applied to a NbSe$_{3}$ nanowire, 
while $P_{\mathrm{SAW}}$ is in an indirect manner. 
Due to this difference, $P_{\mathrm{SAW}}$ is about 1000 
times larger than $P_{\mathrm{ac}}$.
We have evaluated the heating effects both for the SAW 
and ac devices from the temperature dependence 
of resistance of NbSe$_{3}$ nanowire (see Fig.~S5 in~\cite{supplement}). 
The SAW and ac devices are heated up by \SI{2.5}{K} 
at $P_{\mathrm{SAW}} = \SI{7}{mW}$ and by \SI{1}{K} 
at $P_{\mathrm{ac}} = \SI{10}{\uW}$, respectively. 
We also note that the measurements have been conducted at 45--\SI{50}{K}, 
well below $T_{2}$. Thus, the impact of the heating effect should not be large 
enough to affect the CDW sliding. 
There is still a possibility that a piezoelectric field 
in association with SAW is generated on a piezoelectric substrate only at 
the resonant frequency. 
To meet this issue, 
we have calculated the ac voltage in the NbSe$_3$ device $V_{\mathrm{piezo}}$
due to the piezoelectric field accompanied with the SAW.
It turned out that $V_{\mathrm{piezo}}$ for the SAW irradiation is 
10 times smaller than $V_{\mathrm{ac}}$ for the direct application 
of rf current~\cite{supplement}. 
This fact indicates that the amplitude of the piezoelectric field is not enough
to induce the Shapiro steps.

As demonstrated in Fig.~\ref{figure2}, 
the effect of the electromagnetic waves is negligibly small at off-resonant frequencies, 
while both the SAW and electromagnetic wave coexist at the resonant frequency. 
The combination of SAWs and electromagnetic waves might 
induce the present Shapiro steps. 
We have performed a pulsed SAW excitation where 
the SAW is decoupled with the electromagnetic wave and found that 
such a combined effect also does not contribute to the excitation of 
the Shapiro steps,
as detailed in Fig.~S9 in the Supplemental Material~\cite{supplement}. 
Thus, the remaining possibility is the strain induced by the SAW. 
In the present $P_{\mathrm{SAW}}$ range ($\sim \SI{10}{mW}$), 
the longitudinal strain $\varepsilon _{xx}$ applied 
to the NbSe$_3$ nanowire is of the order of $10^{-4}$, 
as detailed in Ref.~\cite{supplement}.
Mori and Maekawa as well as Funami and Aoyama have theoretically addressed  
Shapiro steps in CDW induced 
by sound waves~\cite{Mori_APL_2023, Funami_PRB_2023}. 
However, their theoretical results are different from 
our experimental results. 
One possible reason would be that in their theoretical models, 
the effect of the strain has not been taken into account. 
Nevertheless, the present experimental results 
clearly demonstrate the importance of the interaction between CDW 
and SAWs.

In summary, we have observed Shapiro steps in the CDW state of NbSe$_{3}$ nanowires 
on a piezoelectric substrate induced by SAWs. 
The step width shows an oscillatory behavior as in the case of the conventional 
setup where the ac electric field is directly applied to NbSe$_{3}$ nanowire. 
However, it is drastically reduced with increasing the SAW power and almost 
disappears at high SAW powers. 
By excluding possibilities of electromagnetic waves from IDTs and 
electric fields induced by SAWs, 
we conclude that the strain by the SAW induces the Shapiro steps. 
The present results not only open the door for strain-induced SAW devices
but also shed light upon studies of the dynamics in numerous 
2D materials exhibiting CDWs such as transition metal dichalcogenides and 
rare-earth tritellurides.
It would also be interesting to apply 
the SAW induced strain to Moir\'{e} 
superlattice systems~\cite{Koshino_PRB_2022}. 
At the moment, it is technically difficult to prepare Moir\'{e} 
superlattice systems where the lattice constant is comparable to 
the wavelength of the SAW. Once this is achieved, not only the band 
structure but also phonon modes in Moir\'{e} superlattice could be 
controlled by the strain induced by SAWs.

We thank K.~Aoyama, Y.~Funami, S.~Maekawa, M.~Mori, 
H.~Fukuyama, H.~Matsukawa, M.~Hayashi, and M.~Koshino  
for the fruitful discussions.
This work was supported by JSPS KAKENHI (Grant 
Nos. JP23H00257, JP22KJ2180, JP22J20076), 
JST FOREST (Grant No. JPMJFR2134), and 
the Cooperative Research Project of RIEC, Tohoku University.

\providecommand{\noopsort}[1]{}\providecommand{\singleletter}[1]{#1}%

\end{document}